\documentclass[twocolumn,aps,prl,superscriptaddress,citeautoscript]{revtex4}
\usepackage{graphicx}

\begin{document}

\author{Lan Luan}
\author{Ophir M. Auslaender}
\altaffiliation{Current address: Physics Department, Technion-Israel Institute of Technology, Haifa 32000, Israel}
\author{Thomas M. Lippman}
\author{Clifford W. Hicks}
\author{Beena Kalisky}
\author{Jiun-Haw Chu}
\author{James G. Analytis}
\author{Ian R. Fisher}
\author{John R. Kirtley}
\author{Kathryn A. Moler}
\email[Corresponding author: ]{kmoler@stanford.edu}
\affiliation{Geballe Laboratory for Advanced Materials, Stanford University, Stanford, CA 94305 and Stanford Institute for Materials and Energy Science, SLAC National Accelerator Laboratory, 2575 Sand Hill Road, Menlo Park, CA 94025}

\title{Local measurement of the penetration depth in the pnictide superconductor Ba(Fe$_{0.95}$Co$_{0.05}$)$_2$As$_2$ }

\begin{abstract}
We use magnetic force microscopy (MFM) to measure the local penetration depth $\lambda$ in Ba(Fe$_{0.95}$Co$_{0.05}$)$_2$As$_2$ single crystals and use scanning SQUID susceptometry to measure its temperature variation down to 0.4~K. We observe that superfluid density $\rho_s$ over the full temperature range is well described by a clean two-band fully gapped model. We demonstrate that MFM can measure the important and hard-to-determine absolute value of $\lambda$, as well as obtain its temperature dependence and spatial homogeneity. We find $\rho_s$ to be uniform on the submicron scale despite the highly disordered vortex pinning.
\end{abstract}


\maketitle

The magnetic penetration depth $\lambda$, one of the two fundamental length scales in superconductors \cite{tinkham_introduction_1975}, characterizes many fundamental properties. It evaluates the phase stiffness of the superconducting state by the temperature $T_{\theta}^{max} \propto 1/\lambda^2$ at which phase order would disappear\cite{emery_importance_1995}. It also determines the superfluid density $\rho_s=1/\lambda^2$, the number of electrons in the superconducting phase. However, its absolute value is notoriously difficult to measure, especially in samples that may have either intrinsic or extrinsic inhomogeneity. In this letter, we will report a new technique to measure $\lambda$ by magnetic force microscopy (MFM). The advantage of using local probes over bulk techniques is that it allows us to study the sample homogeneity. We implement this technique to determine $\rho_s$ in a iron-pnictide superconductor Ba(Fe$_{0.95}$Co$_{0.05}$)$_2$As$_2$.

Iron-pnictides superconductors have been under extensive study since their recent discovery \cite{kamihara_iron-based_2008}. The high transition temperature \cite{chen_superconductivity_2008}, the proximity to a magnetic state \cite{ni_effects_2008, chu_determination_2009, zhao_structural_2008}, and the existence of multiple conducting bands \cite{ding_observation_2008,li_probingsuperconducting_2008} combined to make it difficult and interesting to resolve key issues like the superconducting order parameter (OP) symmetry \cite{mazin_unconventional_2008,seo_pairing_2008}, the pairing mechanism \cite{liu_large_2009} and the role of impurities and inhomogeneity \cite{mishra_lifting_2009}. Those problems can be studied by measuring $\rho_s$. When the gap has nodes, $\rho_s(T)$ varies as a power law in $T$ at low $T$, as demonstrated in YB$_2$Cu$_3$O$_{7-\delta}$ \cite{annett_interpretation_1991, hardy_precision_1993}, while a fully gapped OP gives a low-T exponential dependence \cite{prozorov_magnetic_2006}. Since it is difficult to determine $\lambda$, its temperature variation $\Delta\lambda(T)\equiv\lambda(T)-\lambda(0)$ is often measured, which follows the same temperature dependence as $\rho_s$ at low T. Sometimes this approach is sufficient, e.g. linear $\Delta\lambda$ in clean LaFePO over a wide temperature range provides strong evidence of well formed line nodes \cite{fletcher_evidence_2009, hicks_evidence_2009}. However, in the Ba-$122$ family, a steep power-law $\Delta\lambda$ was obtained in the Co-doped compounds \cite{martin_nonexponential_2009, gordon_london_2009} while an exponential $\rho_s$ was measured in the K-doped materials \cite{hashimoto_microwave_2009}. The question waiting for clarification is whether different dopants lead to different OP structure. $\Delta\lambda$ measurement can not infer OP symmetry except for $T\ll T_c$, but for multi-band pnictides, the low-T regime may be dominated by the small-gap regions of the Fermi surface and may be altered by interband impurity scattering \cite{vorontsov_superfluid_2009}. It is thus important to measure the absolute value of $\lambda$ to determine $\rho_s$ over the full temperature range.
\begin{figure}[b]
\includegraphics[width=3.3in]{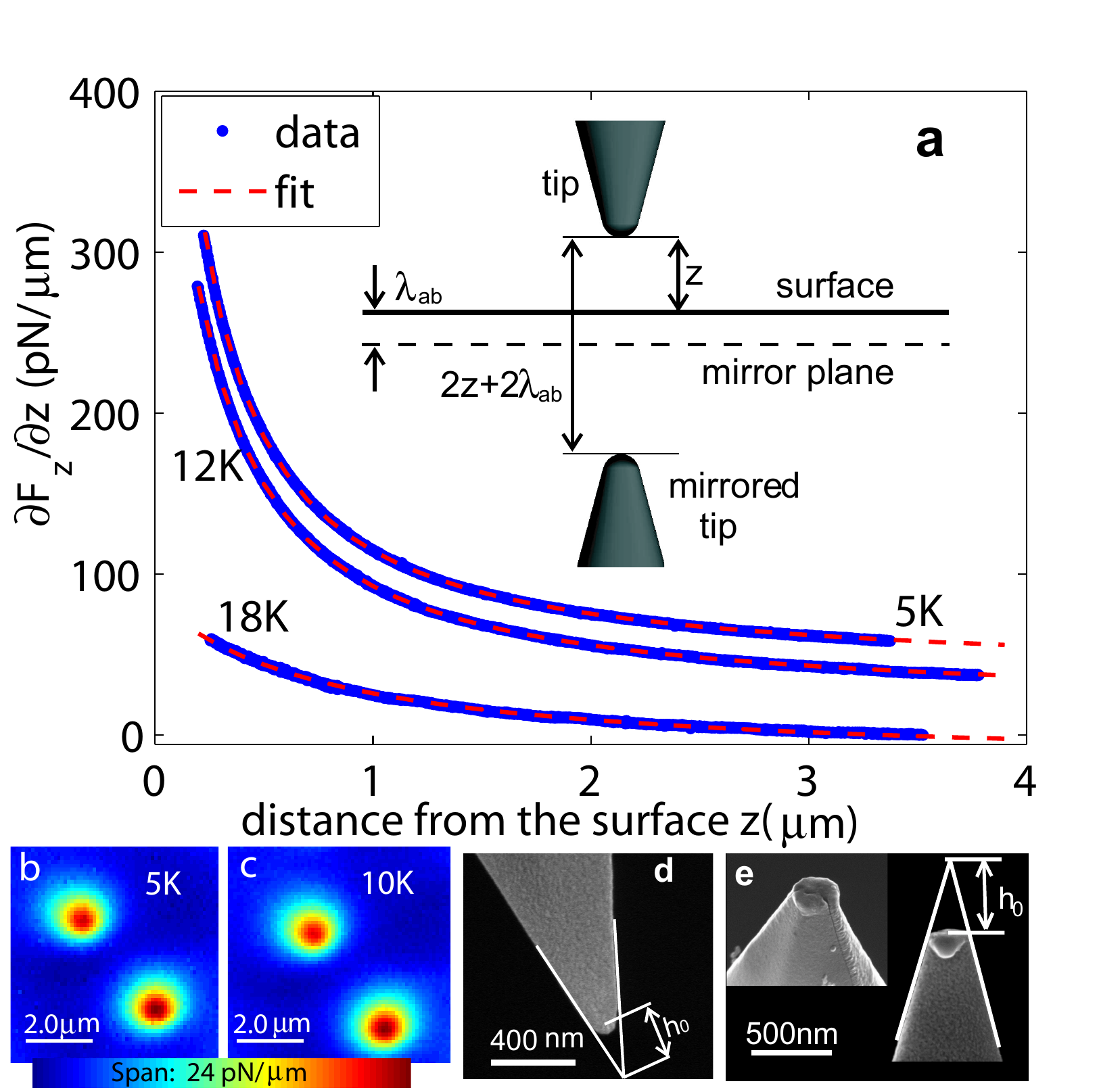}
\caption{\label{fig_twoApproach} Technique to measure $\lambda$ and $\Delta\lambda$ by MFM from Meissner repulsion (a) and vortex imaging (b,c). %
{\bf(a)} $z$ dependence of $\partial F_z/\partial z$ (blue symbols) at $T=5$, 12 and 18 K and the fit to the truncated cone model (red dashed line). {\bf(Inset)}: Sketch to illustrate that the tip-superconductor interaction in the Meissner state can be approximated by the interaction between the tip and its image mirrored through a plane (dashed line) $\lambda_{ab}$ below the surface of the superconductor (solid line) when $z\gg\lambda_{ab}$. Comparing the curves provides $\Delta\lambda_{ab}$ independently of the tip model. Fits give $\lambda_{ab}(T)$ at $T=5$, 12, 18 K to be 0.33, 0.37, $1.10~\mu$m. %
{\bf(b, c)} Images of two vortices ($z=400$ nm) at $5$ K (b) and  $10$ K  (c). The shapes and amplitudes depend on both the magnetic field from vortices and the tip structure, but the similarity shows that both the spatial variation and the temperature-induced change of $\lambda_{ab}$ are small. %
{\bf(d,e)} Scanning electron microscopy images of the tip before (d) and after (e) the measurements. Also shown are the truncation distance $h_0=300\pm30$ nm in (d) and $h_0=400\pm20$ in (e). An accidental crash during the measurement changes the truncation distance $h_0$ from $300\pm30$ nm (d) to $400\pm20$ nm (e). Despite the crash, $\partial F_z/\partial z$ curves taken before and after the crash give the same $\lambda_{ab}(5\rm{K})$ to within $10$ nm.}
\end{figure}

In this paper, we measure the local $\Delta\lambda_{ab}(T)$ and $\lambda_{ab}(T)$, the penetration depth for screening currents flowing in the a-b planes, in electron-doped Ba(Fe$_{1-x}$Co$_{x}$)$_2$As$_2$  single crystals ($x\approx0.05$, $T_c=18.5$ K, grown from self-flux \cite{chu_determination_2009}) from $T=5$ K to $T_c$ by magnetic force microscopy (MFM) [Fig.~\ref{fig_twoApproach}]. We also use scanning SQUID susceptometry (SSS) \cite{tafuri_magnetic_2004} to measure $\Delta\lambda_{ab}(T)$ down to $0.4$ K. We find that $\rho_{s}$ can be well described by a two-band fully gapped OP over the full temperature range. We also use MFM to image and manipulate vortices to measure the homogeneity of $\lambda_{ab}(T)$ and the flux pinning force. We find that $\rho_s$ is uniform to within 10\% or better, although vortex pinning is highly inhomogeneous.

\begin{figure}[pb]
\includegraphics[width=3.3in]{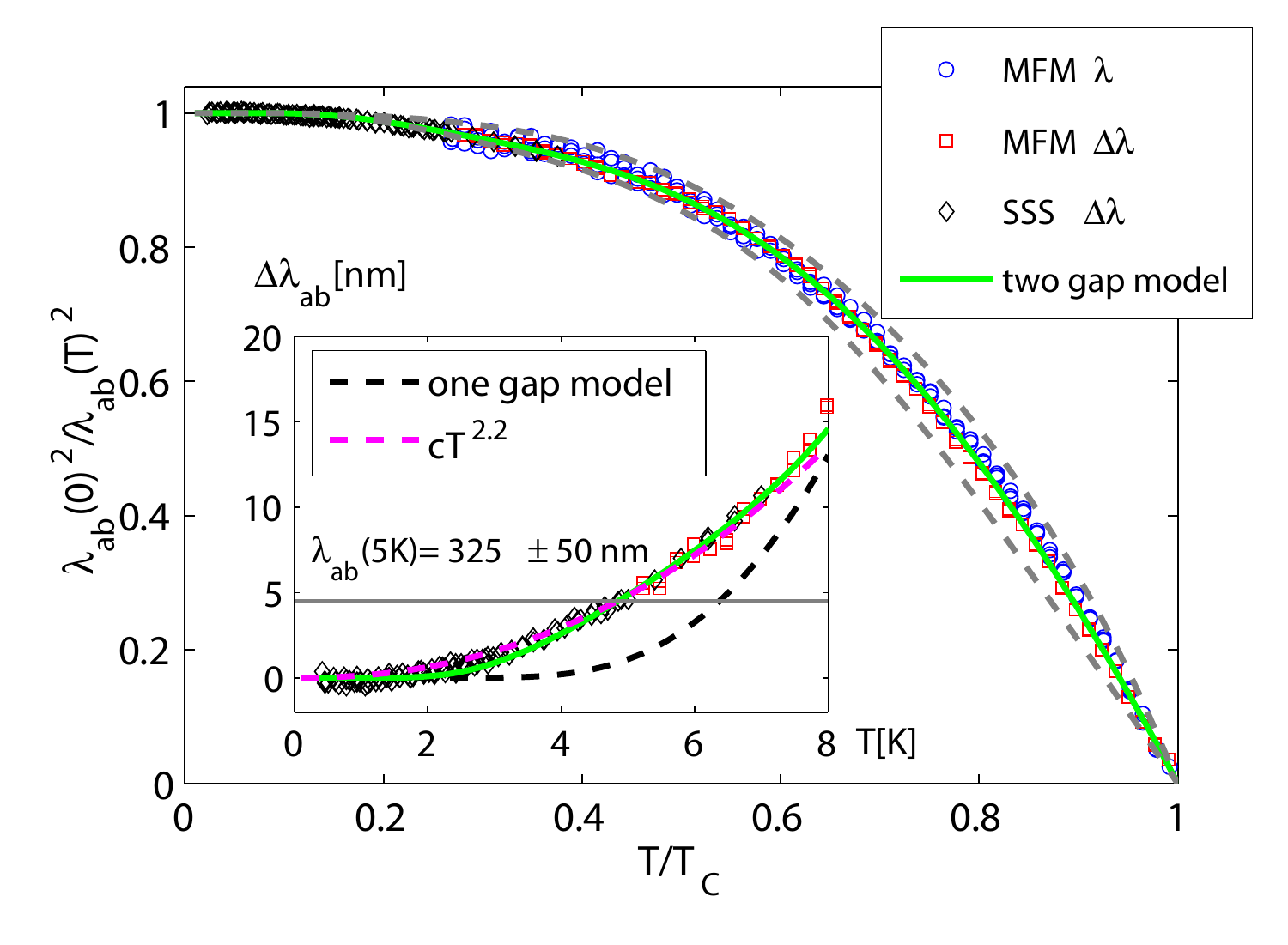}
\caption{\label{fig_penetration} Normalized superfluid density $\rho_s(T)/\rho_s(0)\equiv\lambda_{ab}(0)^2/\lambda_{ab}(T)^2$ vs. T. We determine $\Delta\lambda_{ab}(T)$ by MFM (squares) and by SSS (diamonds) from measuring the change in the diamagnetic response at fixed height. These values are offset to match the absolute value of $\lambda_{ab}(T)$ obtained by fitting the MFM data to the truncated cone model (circles). The green solid line shows a fit of the two-band s-wave model discussed in the main text ($\Delta_1=2.6 T_c$, $\Delta_2=0.8 T_c$, $x=0.88$ and $a=1.4$). The width of the dashed band reflects the uncertainty in $\lambda_{ab}(0)$. %
{\bf Inset:} $\Delta\lambda_{ab}$ vs. $T$ at low $T$.  Black dashed line: one-gap s-wave model with $a=1.5$ and $\Delta_0=1.95 T_c$.  Magenta dashed line: $\Delta\lambda_{ab}(T)=cT^{2.2}$ ( $c=0.14$nm/K$^{2.2}$).}
\end{figure}

In our MFM, a sharp magnetic tip at the end of a flexible cantilever faces the crystal surface, which is parallel to the a-b plane. By measuring the shift in the cantilever's resonant frequency \cite{Albrecht_1991}, we determine $\partial F_z/\partial z$ \cite{straver_controlled_2008}, where $F$ is the force between the tip and the sample, and  $\hat{z}$ is along the tip magnetization direction and is normal to the cantilever and to the crystal a-b surface. $\partial F_z / \partial z$ changes abruptly within a few nanometers of the surface, allowing precise determination of the tip-sample separation $z$. In the Meissner state, the tip-superconductor interaction can be approximated by the magnetic interaction between the tip and its image mirrored through a plane at $z=-\lambda_{ab}$ (Fig.~\ref{fig_twoApproach}a inset) \cite{xu_magnetic_1995}. This local levitation force is determined uniquely by $z+\lambda_{ab}(T)$ for $z\gg\lambda_{ab}$ ($\lambda_c$ does not enter for any source field above a smooth, infinite ab surface) \cite{kogan_meissner_2003}. Thus, changing $T$ at constant $z$ offsets a $\partial F_z / \partial z$ curve along the $\hat{z}$-axis by $\Delta\lambda_{ab}(T)$. To acquire the data labeled as MFM $\Delta\lambda$ in Fig.~\ref{fig_penetration}, we park the tip at $z=500$ nm, change $T$ and acquire $\partial F_z / \partial z$. The $z$ offset required to match $\partial F_z(T) / \partial z$ with a reference curve at $T=5$ K gives $\lambda_{ab}(T)-\lambda_{ab}(5{\rm K})$. Using a similar method for data acquired by SSS in a $^3$He refrigerator \cite{hicks_evidence_2009}, we extend measurements of $\Delta\lambda$ down to $0.4$ K on two nominally identical samples. The SSS results match the MFM results over the common temperature range. By using local scanning probes, we reduce the influence of the complex topography around the sample edges \cite{kogan_meissner_2003}.

Figure~\ref{fig_penetration} shows that $\Delta\lambda_{ab}(T)$ increases very slowly with $T$ at low $T$, inconsistent with the the linear dependence that would be expected for line-nodes. The same behavior appears at three different locations on two samples with SSS and at four different locations with MFM on a third sample. Between $T=0.02T_c$ and $0.4T_c$ $\Delta\lambda_{ab}(T)$ varies by about an order of magnitude less than has been reported for a similar sample using a bulk technique \cite{gordon_london_2009}. At low $T$, $\Delta\lambda_{ab}(T)$ can be described by either a two-band fully-gapped model or by a power law with a small coefficient as described below.

We also extract $\lambda_{ab}(T)$ by modeling the tip-superconductor interaction, with the magnetic tip as a sharp, single domain cone, truncated a at distance $h_0=400\pm20$ nm  from its apex as shown in  Fig.~\ref{fig_twoApproach}e. Within the model, the $z$-dependence of $\partial F_z/\partial z$ is given by:
\begin{eqnarray}\label{eq_1}
\partial F_z(z,T)/ \partial z&-&\left.\partial F_z(z,T) / \partial z\right|_{z=\infty}  =\\ \nonumber &&\hspace{-1in}
A\left(\frac{1}{z+\lambda_{ab}(T)}+\frac{h_0}{\left(z+\lambda_{ab}(T)\right)^2}+\frac{h_0^2}{2\left(z+\lambda_{ab}(T)\right)^3}\right)
\end{eqnarray}
where $A$ is determined by the tip shape and the coating.  The value $A=78 \rm{pN}$ from fitting at $T<T_c/2$ is consistent to within 30\% with the magnetic moment expected from the nominal iron coating on the tip, and with that inferred from the tip-vortex interaction \cite{straver_controlled_2008}. We record $\partial F_z/\partial z$ as a function of $z$ and $T$ and extract $\lambda_{ab}$ at many temperatures by fitting to Eq.~\ref{eq_1} with $A$ and $h_0$ fixed and $\lambda_{ab}$ and $\partial F_z / \partial z (\infty,T)$  allowed to vary separately for each $T$. The fit works well for all $T$ (Fig.~\ref{fig_twoApproach}). The resulting values of $\lambda_{ab}(T)$ are shown in Figure~\ref{fig_penetration} with label "MFM $\lambda$" and  agrees well with the model-independent $\Delta\lambda$. If we consider only statistical errors, we obtain $\lambda_{ab}(5\rm{K})=325\pm5$ nm with 70\% confidence interval. However, the systematic error from the finite width corrections of the tip-geometry is 5\%. In addition, the $\pm 20$ nm uncertainty on $h_0$ leads to $74~\rm{pN}\le A \le 81~\rm{pN}$ by bootstrapping. The extremals of $A$ and $h_0$ gives $\pm 35$ nm systematic error on $\lambda_{ab}(5\rm{K})$. Thus, adding the two main sources of systematic error, we find $\lambda_{ab}(5\rm{K})=325\pm50$ nm.

Knowing $\lambda_{ab}(T)$ gives $\rho_s$ over the full temperature range (Fig.~\ref{fig_penetration}). The fact that $\rho_{s}$ does not saturate at low $T$ is inconsistent with a single-band isotropic gap. A two-band fully gapped OP, which was proposed theoretically \cite{mazin_unconventional_2008,seo_pairing_2008} and tested experimentally \cite{ding_observation_2008,hashimoto_microwave_2009}, describes the data well (Fig.~\ref{fig_penetration}). In the model, $\rho_s(T)=x\rho_1(T)+(1-x)\rho_2(T)$: $\rho_{1,2}(T)$ are the superfluid densities in bands $i=1,~2$, with gaps $\Delta_i(T)=\Delta_i(0)\tanh\left(\frac{\pi T_C}{\Delta_i(0)}\sqrt{a_i\left(\frac{T_C}{T}-1\right)}\right)$; $a_i$ describes the rate of  $\Delta_i(T)$ increasing upon cooling from $T_c$ \cite{prozorov_magnetic_2006}. Our fit (taking into account the systematic error on $\lambda_{ab}(5\rm{K})$) gives $\Delta_1(0)=2.5\pm 0.3T_c$, $\Delta_2(0)=0.70\pm 0.1 T_c$, $x=0.89 \pm 0.06$  and $a_1=1.45 \pm 0.4$ with $a_2\equiv1$. The value of $a_1$ suggests that pairing is likely to be more complicated than phonon-mediated weak coupling \cite{liu_large_2009, stanev_spin_2008}, which would give $a=1$. The magnitude of $\Delta_{1,2}(0)$ is consistent with the scaled down values deduced from optical spectroscopy on similar materials with higher $T_c$ \cite{ding_observation_2008, li_probingsuperconducting_2008}. At low T a power law $cT^n$ where $n=2.2$ and $c=0.14$~nm/K$^{2.2}$ also fits the data. The dominant sources of errors are the calibration accuracy of the scanner, thermal drift, and the breakdown of the assumption of $z\gg\lambda_{ab}$, which together would bound $c$ between $0.12$ and $0.18$~nm/K$^{2.2}$. The small coefficient is inconsistent with that previously reported \cite{martin_nonexponential_2009}. We rule out a nodal OP model since the impurity scattering rate required for such a model \cite{hirschfeld_effect_1993}  to match our data is much higher than that reported in previous works on d-wave cuprates with deliberately added impurities \cite{bonn_comparison_1994, salluzzo_role_2000}. Instead, we interpret this weakened exponential behaviour of $\Delta\lambda_{ab}$ and $\rho_s(T)$ from $0.4$ K all the way to $T_c$ as strong evidence for two full gaps, consistent with the extended s-wave OP \cite{mazin_unconventional_2008,parish_experimental_2008}.

We repeated the touchdown measurement at four positions separated by around $10~\mu$m and obtained $\lambda_{ab}(T=5~\rm{K})=325$ nm, $330$ nm, $325$ nm and $330$ nm. This result suggests that $\lambda_{ab}$ is uniform across the sample.

A second test of uniformity is afforded by measuring the local $T_c$ by mapping the lowest $T$  at which we cannot detect Meissner levitation by MFM (sensitivity corresponds to $\lambda_{ab}(T)>3~\mu$m) or diamagnetic response by SSS (sensitivity corresponds to $\lambda_{ab}(T)>20~\mu$m \cite{kalisky_enhanced_2009}). We find the variation of $T_c$ to be less than $0.5$ K throughout the range of $10\times10 \mu\rm{m}^2$ by MFM and $200\times200 \mu\rm{m}^2$ by SSS.

\begin{figure}[pb]
\includegraphics[width=3.3in]{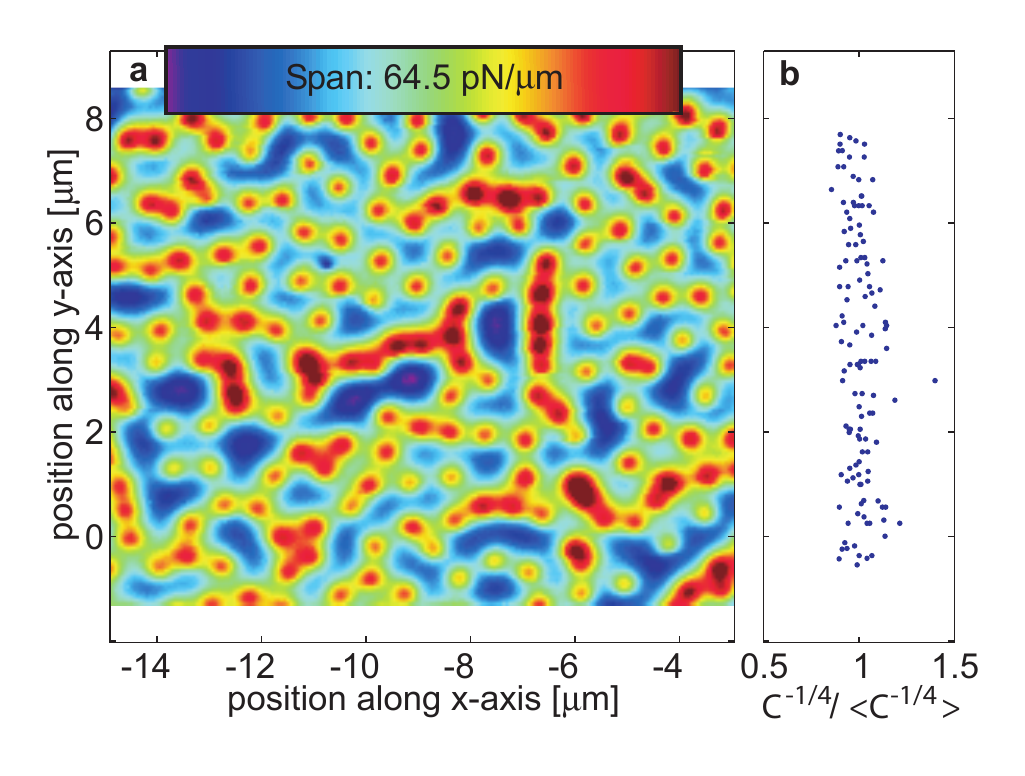}
\caption{\label{fig_uniformity} Spatial uniformity of $\lambda_{ab}$ from vortex imaging at 5 K. {\bf (a)} Image of vortices at $T=5$K, $z=125$ nm and a vortex density giving 3.5 mT. {\bf (b)} The normalized length-scale associated with each vortex peak in (a): ${\cal C}^{-1/4}/<{\cal C}^{-1/4}>$ ($<>$ denotes the mean). We find ${\cal C}^{-1/4}=250(1\pm0.08)$ nm (70\% confidence interval).}
\end{figure}

Vortex imaging provides a third test of $\rho_s(T)$ uniformity. To this end, we cool the sample in an external magnetic field and scan the tip at a constant height $z$ above the surface at 5 K. All vortices appear very similar (Fig.~\ref{fig_uniformity}a), indicating that the spatial variation of $\lambda_{ab}$ is limited. The convolution of the  tip and the vortex field makes it difficult to extract $\lambda_{ab}$ from the vortex imaging. Instead, we calculate the normalized curvature at each vortex peak to quantify the spatial variation: ${\cal C}\equiv\max\left(\partial F_z/\partial z\right)^{-2}\det\left(\frac{\partial^2(\partial F_z/\partial z)}{\partial x_i \partial x_j}\right)$ ($i$, $j$ run over 1, 2 and $x_1\equiv x$, $x_2\equiv y$). The length-scale, ${\cal C}^{-1/4}$, characterizes the spatial extent of the magnetic field from each vortex (Fig.~\ref{fig_uniformity}a). The scatter (Fig.~\ref{fig_uniformity}b) of the normalized ${\cal C}^{-1/4}$ ($\pm$8\%) at constant $z$  gives an estimate for the spatial variation of $\lambda_{ab}$.

\begin{figure}[pbt]
\includegraphics[width=3.3in]{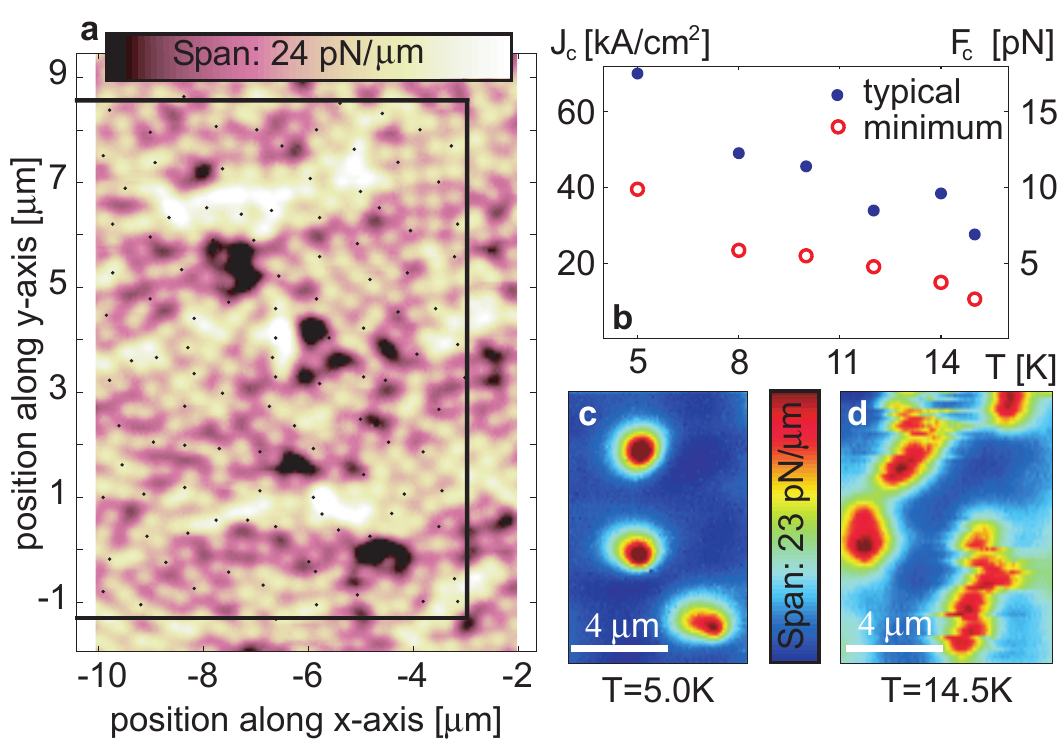}
\caption{\label{fig_pinning}Inhomogeneous vortex pinning.
{\bf(a)} Image of vortices at $T=5$~K, $z=80$~nm and $B=9.5$~mT, overlaid by the vortex positions (dots) in Fig.~\ref{fig_uniformity}a and the boundary of that scan (black frame). The vortex configuration is highly disordered. Vortices avoid the same regions in both scans, taken days apart and many thermal cycles apart. {\bf (b)} Local critical current (left ordinate) and the depinning force (right ordinate) vs. $T$. The comparison of minimum and typical values implies inhomogeneous pinning. {\bf (c)} Image of vortices at $T=5$~K, $z=120$~nm showing that $F_{\rm min}$ only moves the vortex at the bottom. {\bf (d)} Image of moving vortices at $T=14.5$K, $z=430$nm showing that $F_{\rm typ}$ allows us to drag all vortices a distance of several microns.}
\end{figure}

In contrast to the uniform $\rho_{s}$, vortex pinning is very inhomogeneous. Vortices do not form an ordered lattice when field-cooled in fields up to $13$ mT, the highest field that allows us to resolve individual vortices in this material. Instead, vortices always appear in the same regions when we thermal cycle in different fields using different cooling rates (Fig.~\ref{fig_pinning}a). This behavior suggests inhomogeneous pinning. To measure the pinning force distribution, we use the MFM tip to drag individual vortices and to convert the recorded $\partial F_z/\partial z$ to the required force \cite{straver_controlled_2008}. We measure two different forces (Fig.~\ref{fig_pinning}b): the force for dragging the most weakly pinned vortex, $F_{\rm min}$, a measure of the smallest pinning force (Fig.~\ref{fig_pinning}c); and the force for dragging all of the vortices (usually $\lesssim10$) in a field of view, $F_{\rm typ}$, a measures of the typical pinning force (Fig.~\ref{fig_pinning}d). In this sample $2\lesssim F_{\rm typ}/F_{\rm min}\lesssim4$. $F_{\rm typ}\approx 18$~pN at 5K, corresponding to a critical current of $J_c\approx80~{\rm kA/cm}^2$ ($F_c=J_c\Phi_0d$, where $\Phi_0$ is the flux quantum, $d=10 \mu$m is the sample thickness), consistent with the value from bulk measurement of an optimally doped sample \cite{tanatar_anisotropy_2009}. Even at $F_{\rm typ}$, vortices do not follow the tip all the way, indicating the existence of pinning forces larger than $F_{\rm typ}$. In fact, $F_{\rm typ}$ is still at least an order of magnitude smaller than the force required to stabilize vortices in the dense clusters we see (the vortex-vortex interaction for a pair separated by 400~nm corresponds to a current density of $3~{\rm MA/cm}^2$). We do not detect any correlation between pinning and superfluid density, suggesting that strong pinning exists without affecting superconductivity on the scale of $\lambda_{ab}$. The ability to measure the absolute value of the penetration depth despite a disordered vortex configuration is important, since the most commonly used method, muon-spin-rotation \cite{williams_muon_2009}, assumes an ordered vortex configuration.

To conclude, by measuring $\lambda_{ab}(T)$  and $\Delta\lambda_{ab}(T)$ locally we find that underdoped Ba(Fe$_{1-x}$Co$_x$)$_2$As$_2$ ($x\approx0.05$) has homogenous $\rho_s$ whose temperature dependence can be described by a two-band fully-gapped OP. This result provides thermodynamic evidence for fully gapped models such as the proposed extended s-wave model \cite{mazin_unconventional_2008,seo_pairing_2008} for Co-doped $122$ pnictides and shows that it has the similar OP structure as the K-doped, despite the different dopants and substitution cites. We obtain $\lambda_{ab}(0)=325\pm 50$~nm, which gives $T_{\theta}^{max}= A(\hbar c)^2 a/(16\pi e^2 \lambda^2)\approx 260$~K, where $a=\sqrt{\pi}\xi_c$, $\xi_c=1.1$ nm \cite{sun_comparative_2009} and A=2.2 in the three-dimension limit \cite{emery_importance_1995, tanatar_anisotropy_2009}. $T_{\theta}^{max} \gg T_c$, hinting that phase fluctuations are not as important here as in the underdoped cuprates \cite{emery_importance_1995}. Instead, $T_c$ in the underdoped iron-pnictides may be suppressed by the competition with non-superconducting phases. MFM allows us to obtain the superfluid density and to map its spatial variation down to the submicron scale. This capability may be useful to study how different phases compete for charge carriers.

Acknowledgment: This work is supported by the Department of Energy, Office of Basic Energy and Sciences under contract DE-AC02-76SF00515. We thank A.~Bernevig, A.~Chubukov, S.~Kivelson, D.~Scalapino, O.~Vafek and A.~Vishwanath for helpful discussions.

\end{document}